# Differential diagnosis and molecular stratification of gastrointestinal stromal tumors on CT images using a radiomics approach

Martijn P.A. Starmans*[1,2] MSc, Milea J.M. Timbergen*[3,4] MD, Melissa Vos[3,4] MD, Michel Renckens[1] MD, Dirk J. Grünhagen[3] MD PhD, Geert J.L.H. van Leenders[5] MD PhD, Roy S. Dwarkasing[1] MD PhD, François E. J. A. Willemssen[1] MD PhD, Wiro J. Niessen [1,2,6] PhD, Cornelis Verhoef[3] MD PhD, Stefan Sleijfer[4] MD PhD, Jacob J. Visser[1] MD PhD, and Stefan Klein [1,2] PhD

[1] Department of Radiology and Nuclear Medicine, Erasmus Medical Center, Rotterdam, the Netherlands

[2] Department of Medical Informatics, Erasmus Medical Center, Rotterdam, the Netherlands

[3] Department of Surgical Oncology, Erasmus MC Cancer Institute, Erasmus Medical Center, Rotterdam, the Netherlands

[4] Department of Medical Oncology, Erasmus MC Cancer Institute, Erasmus Medical Center, Rotterdam, the Netherlands

[5] Department of Pathology, Erasmus Medical Center, Rotterdam, the Netherlands

[6] Faculty of Applied Sciences, Delft University of Technology, Delft, the Netherlands

*These authors contributed equally

**Abstract**

Distinguishing gastrointestinal stromal tumors (GISTs) from other intra-abdominal tumors and GISTs molecular analysis is necessary for treatment planning, but challenging due to its rarity. The aim of this study was to evaluate radiomics for distinguishing GISTs from other intra-abdominal tumors, and in GISTs, predict the *c-KIT, PDGFRA, BRAF* mutational status and mitotic index (MI). All 247 included patients (125 GISTS, 122 non-GISTs) underwent a contrast-enhanced venous phase CT. The GIST vs. non-GIST radiomics model, including imaging, age, sex and location, had a mean area under the curve (AUC) of 0.82. Three radiologists had an AUC of 0.69, 0.76, and 0.84, respectively. The radiomics model had an AUC of 0.52 for *c-KIT*, 0.56 for *c-KIT exon 11*, and 0.52 for the MI. Hence, our radiomics model was able to distinguish GIST from non-GISTS with a performance similar to three radiologists, but was not able to predict the *c-KIT* mutation or MI.

**Introduction**

Gastrointestinal stromal tumors (GISTs) are rare mesenchymal tumors of the gastrointestinal tract, with an estimated incidence between 10-15 cases per million inhabitants per year (Soreide et al., 2016; Verschoor et al., 2018). The most common tumor locations are the stomach (56%) and the small intestine (32%); less common locations are the esophagus (<1%) and the colorectal region (6%) (Soreide et al., 2016). Differentiating GISTs from other intra-abdominal tumors (non-GISTs), such as schwannomas, leiomyosarcomas, leiomyomas, esophageal/gastric junctional adenocarcinomas, and lymphomas is highly important for treatment planning (Miettinen & Lasota, 2006b). Computed tomography (CT) is the imaging modality of choice in GIST diagnosis (Lau et al., 2004), but as the differential diagnosis remains challenging, assessment through an invasive tissue biopsy is generally required (Demetri et al., 2010). A non-invasive and quicker alternative may aid in the early assessment of GISTs.

Treatment planning of GISTs is also based on their molecular profile. The mitotic index (MI) reflects the proliferative rate of GISTs, correlates with survival and risk of metastatic spread (Rudolph, Gloeckner, Parwaresch, Harms, & Schmidt, 1998), and as such determines whether or not a patient with localized disease should get adjuvant systemic treatment. Treatment decisions are also based on the mutational status of GISTs. *PDGFRA* exon 18 mutated (Asp842Val) GISTs are resistant to imatinib (Cassier et al., 2012) and alternative treatments are being explored in this specific subgroup. GISTs with a *c-KIT* exon 11 mutation also have shown a greater sensitivity for imatinib than those with a *c-KIT* exon 9 mutations (Miettinen & Lasota, 2006b), hence the latter are often treated with a higher imatinib dose. The MI and these genetic mutations are currently assessed through an invasive tissue biopsy.

The field of radiomics relates imaging features to molecular characteristics in order to non-invasively contribute to diagnosis, prognosis and treatment decisions. Several radiomics studies have shown promising results in risk stratification of GISTs (Ba-Ssalamah et al., 2013; Chen et al., 2019; Feng et al., 2018; Kurata et al., 2018; Liu et al., 2018; Ning et al., 2019; Xu, Ma, et al., 2018; Yang et al., 2018; Zhou et al., 2016; Zhuo, Li, & Zhou, 2018). However, radiomics has not been previously used to distinguish GISTs from non-GISTs, nor to predict the mutational status or the MI.

The aim of this study was to evaluate whether an automatically optimized radiomics model based on CT is capable of 1) differentiating GISTs from other intra-abdominal tumors resembling GIST prior to treatment, i.e. the differential diagnosis; and 2) predicting the presence and type of mutation (*BRAF, PDGFRA* and *c-KIT)* and the MI of GISTs, i.e. the molecular analysis.

**Methods**

*Data collection*

Approval by the [anonymized] institutional review board was obtained (MEC-2017-1187). Patients from our institute between 2004-2017 with a histopathologically proven primary GIST or intra-abdominal tumors resembling GIST with at least a contrast-enhanced venous phase CT prior to treatment (Kang et al., 2013; Miettinen & Lasota, 2006b), were retrospectively included. Several GISTs may have been included in the Dutch GIST registry. Exact numbers on potential overlap with previous studies using the registry cannot be determined. As no radiomics studies on this registry have been published, potential overlap has little relevance. Age at diagnosis, sex, and tumor location were collected. Tumor location was based on radiology reports and categorized into: (distal) esophagus, stomach, small intestine, colon, rectum, pelvis, mesentery, uterus, and other. The sample sizes of the non-GIST and the GIST cohort were matched. The non-GIST subtypes were balanced, i.e. a similar number of patients per subtype was randomly included.

GISTs with a known mutation status and/or MI, prior to therapy were included for the molecular analysis. Both were obtained from pathology reports and analyzed on either the primary lesion or, in case of metastatic disease at first presentation, on secondary lesions. The mutation was categorized as 'absent' or 'present' for each type (e.g. *c-KIT*) and subtype (e.g. *c-KIT* exon 11). The MI (expressed in high power fields (HPF), magnification 40x, totaling 5mm$^2$), determined on biopsy or excision material, was split into low (≤5/50 HPF) and high (>5/50 HPF) (Miettinen & Lasota, 2006a). An adjusted MI was calculated per 50 HPF when the MI was not counted per 50 HPF. In case of unknown mutation status or MI, the case was excluded from the particular analysis.

*Radiomics*

The radiomics workflow is depicted in **Figure 1**, adapted from (Vos et al., 2019). The tumors were all manually segmented once by one of two clinicians under supervision of a musculoskeletal radiologist (5 years of experience) using in-house developed software (Starmans, Miclea, et al., 2018). A subset of 30 GISTs was segmented by both clinicians, in which intra-observer variability was evaluated through the pairwise Dice Similarity Coefficient (DSC), with DSC > 0.70 indicating good agreement (Zou et al., 2004). For each lesion, 564 features quantifying intensity, shape, and texture were extracted. For details, see Supplemental Material 1. To create a decision model from the features, the WORC toolbox was used (Starmans, Van der Voort, Phil, & Klein, 2018; Starmans et al., 2019; Vos et al., 2019). In WORC, the decision model creation consists of several steps, e.g. feature selection, resampling, and machine learning. WORC performs an automated search amongst a variety of algorithms for each step and determines which combination maximizes the prediction performance on the training set. For details, see Supplemental Material 2. The code for the feature extraction and model creation has been published open-source (Starmans, 2020).

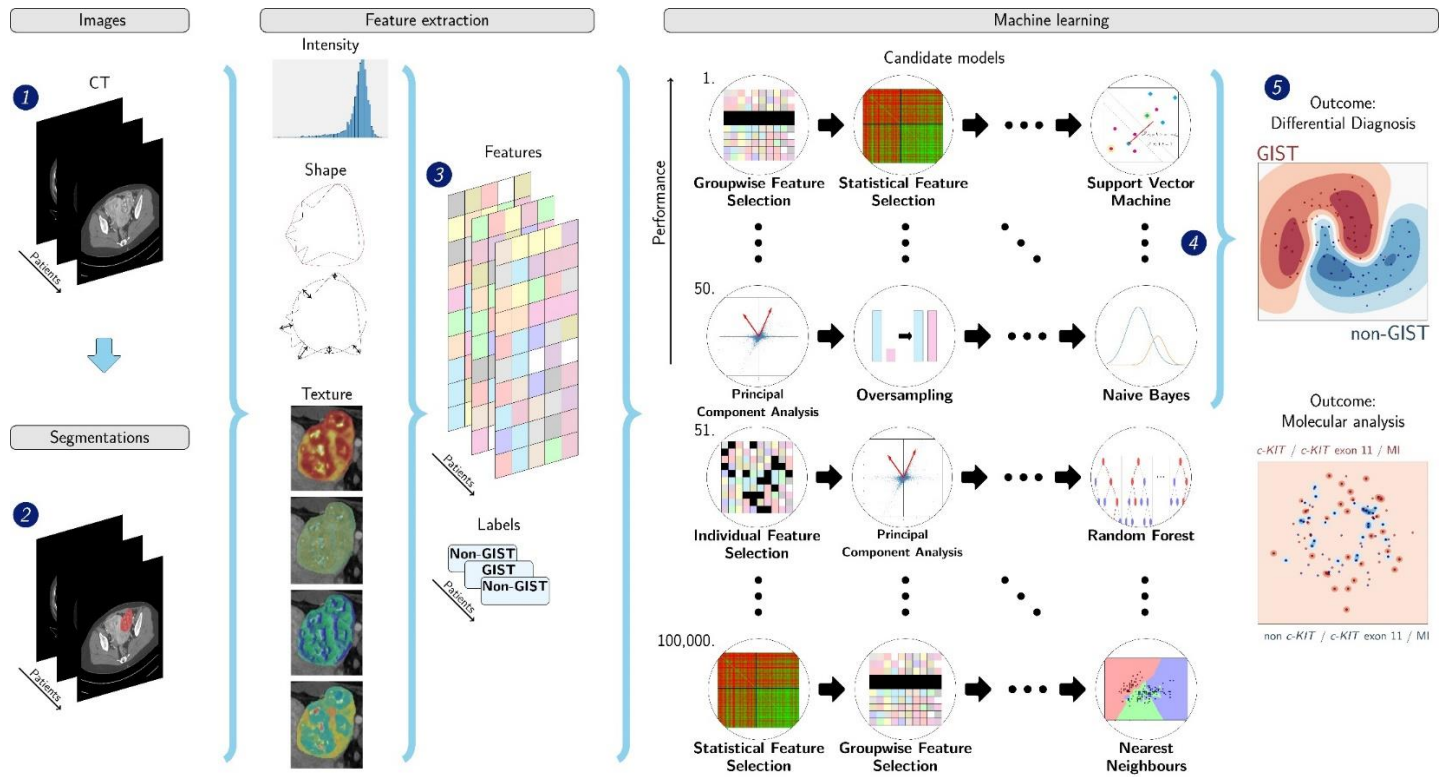

**Figure 1.** Schematic overview of the radiomics approach. Input to the algorithm are the CT images (1). Processing steps then include segmentation of the tumor (2), feature extraction (3) and the creation of machine learning decision models (5), using an ensemble of the best 50 workflows from 100,000 candidate workflows (4), which are different combinations of the different processing and analysis steps (e.g. the classifier used). *Abbreviations: GIST, gastrointestinal stromal tumor.

*Robustness to segmentation and image acquisition variations*

Radiomics' robustness to segmentation variations was assessed using the intra-class correlation coefficient (ICC) of the features on the subset of 30 GISTs which were segmented by two observers. ``Good'' and ``excellent'' reliability were defined by ICC > 0.75 and ICC > 0.90, respectively [26]. Moreover, the impact of ICC-based feature selection on model performance was assessed by creating models using only features with good or excellent reliability.

Robustness to variations in the acquisition parameters was assessed by using ComBat harmonization [27; 28]. In ComBat, feature distributions are harmonized for variations in the imaging acquisition, e.g. due to differences in hospitals, manufacturers, or acquisition parameters. When dividing the dataset into groups based on these variations, the groups have to remain sufficiently large to estimate the harmonization parameters. In our study, groups were defined based on manufacturer alone, or based on protocol, defined as the combination of manufacturer and slice thickness (above or below the median). No moderation variable was used.

*Experimental setup*

Evaluation of all models was done through a 100x random-split cross-validation. In each iteration, the data was randomly split in 80% for training and 20% for testing in a stratified manner, to make sure the distribution of the classes in all sets was similar to the original, see **Supplemental Figure S1**. Within the training set, model optimization was performed using an internal cross-validation (5x). Hence, all optimization was done on the training set to eliminate any risk of overfitting on the test set.

Performance was evaluated using the Area Under the Curve (AUC) of the Receiver Operating Characteristic (ROC) curve, balanced classification accuracy (BCA) (Tharwat, 2018), sensitivity, and specificity. The positive classes were defined as: GIST, the presence of the mutations, and a high MI in the respective analyses. The 95% confidence intervals (CIs) were constructed using the corrected resampled t-test based on the results from all 100 cross-validation iterations (Nadeau & Bengio, 2003). Both the mean and the confidence intervals are reported. ROC confidence bands were constructed using fixed-width bands (Macskassy, Provost, & Rosset, 2005).

To assess the predictive value of the various features, models were trained based on: 1) volume; 2) location; 3) age and sex; 4) imaging; 5) age, sex, and imaging; and 6) age, sex, imaging, and tumor location. Models 2 and 6, assessing the predictive value of location, were included in the differential diagnosis because the radiologists also used tumor location.

In the mutation stratification, only the subset of patients with a known mutation (sub)type was taken into account for each analysis.

*Model insight*

To explore the predictive value of individual features, the Mann-Whitney *U* univariate statistical test was used for continuous variable, and a Chi-square test for categorical variables. P-values were corrected for multiple testing using the Bonferroni correction, and were considered statistically significant at a p-value <0.05. To gain insight into the models, the patients were ranked based on the consistency of the model predictions. Typical examples for each class consisted of the patients that were correctly classified in all cross-validation iterations; atypical vice versa. To estimate model robustness to segmentation and acquisition protocol variations, for the differential diagnosis, additional imaging-only models (i.e. model 4) were created using only reliable features through ICC-based feature selection and ComBat harmonization, respectively.

*Performance of the radiologists*

To compare the models with clinical practice, three radiologists (5, 15 and 12 years of experience) independently evaluated the tumors. Evaluation was done on a ten-point scale to indicate the scoring certainty, i.e. 1 = strongly disagree GIST, 5 = mildly disagree GIST, 6 = mildly agree GIST, 10 = strongly agree GIST. The radiologists were blinded for the diagnosis but had access to the CT scan, patient age and sex. Only the differential diagnosis was scored, as the mutation and MI are based on pathology in clinical practice. The radiologists' agreement was evaluated using Cohen's Kappa. To enable direct statistical comparison between the radiologists' performances on the one hand and the best radiomics model on the other hand, the radiomics model was evaluated in an additional leave-one-out cross-validation, after which the DeLong test was used to compare the AUCs (DeLong, DeLong, & Clarke-Pearson, 1988).

**Results**

*Study population and dataset*

The dataset included 247 patients (125 GISTs, 122 non-GISTs), which were all included in the differential diagnosis analysis. Sclerosing mesenteritis (N=16) and inflammatory fibroid polyp (N=4) were excluded due to their small numbers. Clinical characteristics of the dataset are summarized in **Table 1**. The dataset of 247 CT scans originated from 66 different scanners, resulting in variation in the acquisition protocols, see **Table 1**. The scans originated from four different manufacturers (Siemens, Berlin, Germany: 126, Philips, Eindhoven, the Netherlands: 63, General Electric, Boston, United States: 10, Toshiba, Tokyo, Japan: 48). On the subset of 30 GISTs that was segmented by both observers, the mean DSC was 0.84 (standard deviation of 0.20), indicating good agreement.

**Table 1.** Clinical and CT scan characteristics of the dataset.

| | GISTs | Schwannoma | Leiomyo-sarcoma | Leiomyoma | Esophageal/ gastric junctional adenocarcinoma | Lymphoma |
|---|---|---|---|---|---|---|
| **Number** | 125 | 22 | 25 | 25 | 25 | 25 |
| **Sex** | | | | | | |
|   Male | 66 (53%) | 11 (50%) | 7 (28%) | 6 (24%) | 16 (64%) | 18 (72%) |
|   Female | 59 (47%) | 11 (50%) | 18 (72%) | 19 (76%) | 9 (36%) | 7 (28%) |
| **Age at diagnosis** [a] | 64 (56-72) | 59 (45-67) | 60 (53-71) | 49 (41-59) | 65 (56-74) | 62 (52-67) |
| **Tumor location** [b] | | | | | | |
|   (Distal) esophagus | - | - | - | 6 (24%) | 5 (20%) | - |
|   Stomach | 80 (64%) | 2 (9.1%) | 1 (4%) | 3 (12%) | 20 (80%) | 2 (8%) |
|   Small intestine | 29 (23%) | - | 1 (4%) | - | - | 4 (16%) |
|   Colon | 1 (1%) | - | 2 (8%) | - | - | 1 (4%) |
|   Rectum | 7 (6%) | - | - | - | - | - |
|   Pelvis | 1 (1%) | 7 (31.8%) | 5 (0%) | 2 (8%) | - | 1 (4%) |
|   Mesentery | - | - | - | - | - | 7 (28%) |
|   Uterus | - | - | 2 (8%) | 13 (52%) | - | - |
|   Other | 7 (6%) | 13 (59.1%) | 14 (56%) | 1 (4%) | - | 10 (40%) |
| **Tumor volume (cl)** [a] | 15.7 (4.3-52.6) | 13.9 (1.6-29.7) | 12.9 (6.7-99.6) | 8.2 (1.6-25.5) | 1.6 (0.7-3.1) | 9.4 (4.6-29.4) |
| **Acquisition protocol** | | | | | | |
|   Slice thickness (mm) [a,c] | 5.0 (3.0-5.0) | 5.0 (2.0-6.0) | 5.0 (3.0-5.0) | 3.0 (3.0-5.0) | 4.0 (3.0-5.0) | 3.0 (3.0-3.0) |
|   Pixel spacing (mm) [a,c] | 0.72 (0.68-0.78) | 0.74 (0.68-0.79) | 0.72 (0.68-0.78) | 0.75 (0.68-0.84) | 0.74 (0.66-0.78) | 0.77 (0.69-0.85) |
|   Tube current (mA) [a,c] | 189 (129-283) | 162 (115-206) | 221 (160-349) | 210 (147-395) | 210 (142-312) | 207 (145-301) |
|   Peak kilovoltage [a,c] | 120 (100-120) | 120 (120-120) | 120 (100-120) | 120 (100-120) | 120 (100-120) | 100 (100-100) |

[a] Median (inter quartile range)
[b] Percentages may not add up to 100% because of rounding
[c] Other values than those given in the median and inter quartile range do occur
* Abbreviations: GIST: gastrointestinal stromal tumor; cl: centiliter; mm: milimeter; mA: mili Ampére

Two patients were excluded for the molecular radiomics analysis as the molecular characteristics were obtained after receiving systemic treatment. A total of 123 GISTs were included in the cohort for the molecular analysis. The mutation analysis was performed on tissue obtained from the primary lesion, except for three patients for which this was performed on a metastatic hepatic lesion. A *c-KIT* mutational analysis was performed in 98/123 (80%) GIST patients. One patient had a *c-KIT* mutation which was not further specified. Twenty-six (27%) patients had no *c-KIT* mutation. The majority of patients had a *c-KIT* exon 11 mutation (N=59, 60%). Due to the low numbers of *c-KIT* exon 9 (N=10), *c-KIT* exon 13 (N=2), *PDGFRA* (N=14), and *BRAF* (N=0), these mutations were excluded from further analysis.

The MI was available in 90/123 (73%) GISTs (55 low, 35 high). The MI of 33 (37%) GISTs was converted to the adjusted MI. The MI was determined on excision material in 54 (60%) patients, and on biopsy material in 36 (40%) patients, including one patient in which the MI was based on the hepatic GIST metastasis.

*Evaluation of models for the differential diagnosis*

The performances of the models distinguishing GISTs from non-GISTs are shown in **Table 2** and **Figure 2**. On average, model 1, based solely on volume, did not perform well (AUC of 0.56). Model 2, based on location, performed better (AUC of 0.82), but showed a sharp cutoff in the ROC curve (**Figure 3b**). Model 3, based on age and sex, did not perform well (AUC of 0.61). Model 4, based on CT imaging features, performed better with a mean AUC of 0.74. Model 5, combining imaging with age and sex, did not yield an improvement (AUC of 0.70). Model 6, adding tumor location, did yield an improvement (AUC of 0.82).

**Table 2.** Performance of the radiomics models for the differential diagnosis based on 1) volume; 2) location; 3) age and sex; 4) imaging features; 5) imaging features, age and sex; and 6) imaging features, age, sex and tumor location, and of the three radiologists (R1, R2 and R3). Values are presented with their 95% confidence intervals

|  | Model 1 Volume | Model 2 Location | Model 3 Age+sex | Model 4 Imaging | Model 5 Imaging+age+sex | Model 6 Imaging+age+sex+location | R1 | R2 | R3 |
|---|---|---|---|---|---|---|---|---|---|
| **AUC** | 0.56 [0.48, 0.64] | 0.82 [0.76, 0.88] | 0.61 [0.55, 0.68] | 0.74 [0.67, 0.81] | 0.70 [0.64, 0.77] | 0.82 [0.76, 0.87] | 0.69 | 0.76 | 0.84 |
| **BCA** | 0.55 [0.49, 0.60] | 0.82 [0.76, 0.88] | 0.56 [0.51, 0.62] | 0.67 [0.60, 0.74] | 0.66 [0.59, 0.73] | 0.74 [0.68, 0.80] | 0.67 | 0.67 | 0.76 |
| **Sensitivity** | 0.28 [0.15, 0.41] | 0.93 [0.88, 0.98] | 0.62 [0.54, 0.71] | 0.58 [0.46, 0.72] | 0.58 [0.46, 0.70] | 0.71 [0.61, 0.82] | 0.74 | 0.90 | 0.78 |
| **Specificity** | 0.82 [0.70, 0.93] | 0.71 [0.60, 0.82] | 0.50 [0.41, 0.60] | 0.75 [0.67, 0.84] | 0.75 [0.64, 0.85] | 0.77 [0.68, 0.85] | 0.60 | 0.44 | 0.74 |

\* Abbreviations: AUC: area under the receiver operating characteristic curve; BCA: balanced classification accuracy; R1, R2 and R3: radiologists 1, 2 and 3

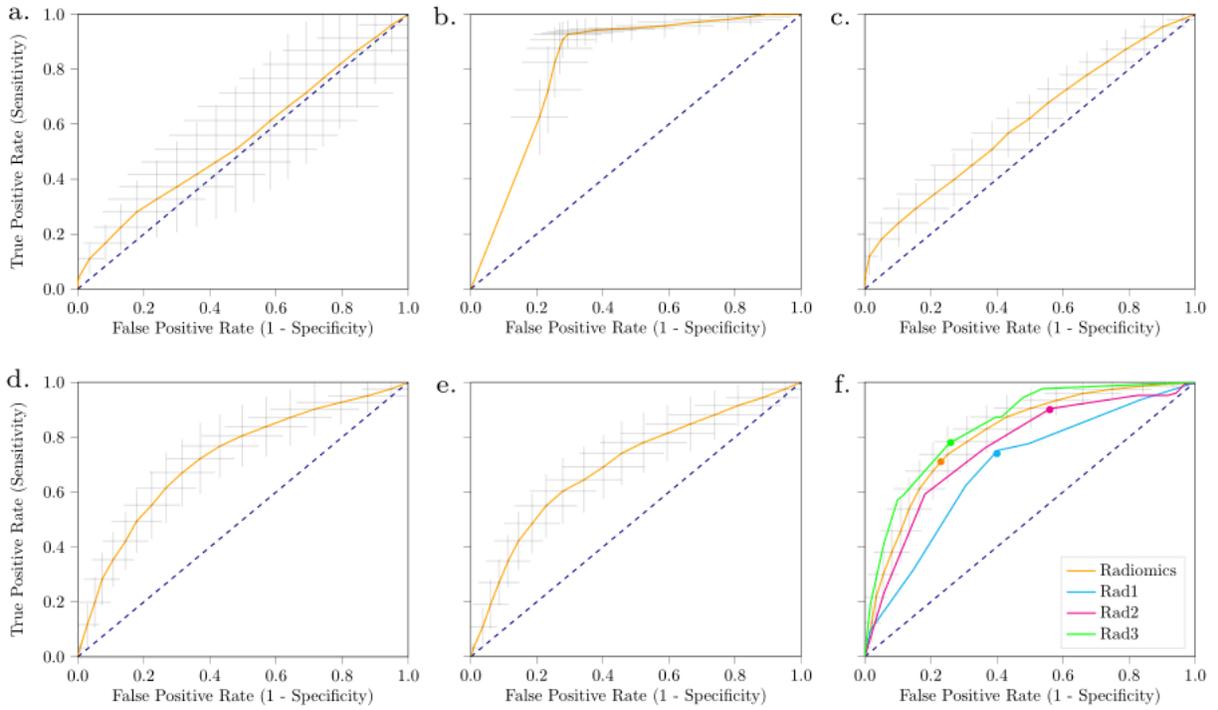

**Figure 2.** Receiver operating characteristic curves of the radiomics models for the differential diagnosis based on volume (**a**); location (**b**); age and sex (**c**); imaging (**d**); imaging, age, and sex (**e**); imaging, age, sex, and tumor location (**f**). Additionally in figure (**f**), the curves for scoring by three radiologists are shown, and the cut-off points for both the radiomics model and the radiologists. For the radiomics models, the grey crosses identify the 95% confidence intervals of the 100x random-split cross-validation; the orange curve is fit through their means.

*Comparison with radiologists*

The performance of the radiologists is shown in **Table 2** and **Figure 2**. Compared to model 6, which had the same inputs, i.e. based on imaging, age, sex, and tumor location, the AUCs of the first two radiologists (0.69 and 0.76) were lower, while the AUC of the third radiologists was similar (0.84). All radiologists had a relatively high sensitivity (0.74, 0.90, and 0.78), but a low specificity (0.60, 0.44, and 0.74). Cohen's kappa measures between the pairs of radiologists were 0.20, 0.31 and 0.33, all indicating poor inter-observer agreement. The Delong test between the pairs of radiologists indicated a significant difference in performance for radiologists 1 versus 3 ($p=6 \times 10^{-5}$) and 2 versus 3 (p=0.01); for radiologist 1 versus 2, the power was too low to claim insignificance. Radiomics model 6 evaluated in a leave-one-out cross-validation (AUC of 0.82) also performed statistically significantly better than the first radiologist (p=0.0018); for comparison with the other radiologists, the power was too low to claim insignificance.

*Evaluation of models for the molecular analysis*

For the c-KIT mutation stratification and MI predictions, the performance of the radiomics model based on age, sex and imaging features (model 5) is depicted in **Table 3**.

**Table 3.** Performance of radiomics model 5, based on imaging, age, and sex, for the GIST mutation stratification and the mitotic index for A) *c-KIT* presence vs. absence B) *c-KIT* exon 11 presence vs. absence; and C) mitotic index (≤5/50 HPF vs. >5/50 HPF**)**. The number of patients included in each analysis (*N*) is mentioned in the heading. Values are presented with their 95% confidence intervals.

|  | Model 5A *c-KIT* (N=98) | Model 5B *c-KIT* exon 11 (N=96) | Model 5C Mitotic index (N=90) |
|---|---|---|---|
| **AUC** | 0.52 [0.38, 0.66] | 0.56 [0.44, 0.67] | 0.52 [0.38, 0.65] |
| **BCA** | 0.49 [0.46, 0.52] | 0.52 [0.44, 0.61] | 0.51 [0.41, 0.60] |
| **Sensitivity** | 0.97 [0.91, >1.00] | 0.78 [0.64, 0.91] | 0.30 [0.12, 0.47] |
| **Specificity** | 0.01 [<0.00, 0.07] | 0.27 [0.11, 0.43] | 0.71 [0.56, 0.87] |

* Abbreviations: AUC: area under the receiver operating characteristic curve; BCA: balanced classification accuracy; PPV: positive predictive value; NPV: negative predictive value

In the mutation stratification, the radiomics models had a mean AUC of 0.52, a low specificity (0.01), and a high sensitivity (0.97) for predicting the presence of a *c-KIT* mutation in general (model 5A). Predicting the presence of a *c-KIT* exon 11 mutation (model 5B) performed similar (AUC of 0.56). The MI prediction (model 5C) had a mean AUC of 0.52, a high specificity (0.71) and a low sensitivity (0.30). All models thus focus on the majority class and perform close to guessing, as is confirmed by the ROC curves in **Supplemental Figure S2**. As models 1, 3 and 4 include a subset of the features from model 5, which already did not perform well, these results are omitted. Model 2 and 6 were only used in the differential diagnosis.

*Model insight*

As the molecular analysis models did not perform well, the model insight analysis was only conducted for the differential diagnosis. The p-values of the feature importance analysis are shown in **Supplemental Table S1.** In total, 43 features had significant p-values after Bonferroni correction ($1.1 \times 10^{-17}$ to $4.6 \times 10^{-2}$). These included the tumor location ($1.1 \times 10^{-17}$), two intensity features, three orientation features, four shape features of which three related to the tumor area, and 33 texture features. A list of these features and their p-values has been added to the mentioned published code (Starmans, 2020). Volume was not found to be significant.

Results on ranking patients from typical to atypical are only shown for the model based on imaging, i.e. model 4, as we were interested in the imaging features that defined typical GISTs. Of the 247 patients, 104 tumors (44 GISTs, 60 non-GISTs, 42%) were always classified correctly, and were thus considered typical. Twenty-nine tumors (18 GISTs, 11 non-GISTs, 12%) were always classified incorrectly and thus atypical. In **Figure 3**, four CT slices of such typical and atypical examples of GISTs are shown. Visual inspection of the tumors defined as typical or atypical by the radiomics model showed a relation with necrosis (more present in typical GIST, typically a necrotic core) and shape (more compact, circular and non-lobulated for typical GIST). The patients which were equally often classified as GIST and non-GIST in the cross-validation iterations were mostly small tumors. These typical characteristics and the difficulty with small tumors correspond to the literature for GIST risk stratification (Maldonado et al., 2017; Zhou et al., 2016). Smaller tumors were also more often misclassified by the radiologists in our study.

A list of the ICC values of all imaging features has been added to the mentioned published code (Starmans, 2020). Of the 564 imaging features, 327 (58 %) had an ICC > 0.75 and thus good reliability, 197 (34%) had an ICC > 0.90 and thus excellent reliability. Only using features with a good or excellent reliability in model 4 did not substantially alter the performance (AUC of 0.76 and 0.75, respectively), see **Supplemental Table S2**. Similarly, using ComBat to harmonize the features for manufacturer or protocol differences did not substantially alter the performance either (AUC of 0.76 and 0.73, respectively), see **Supplemental Table S2**.

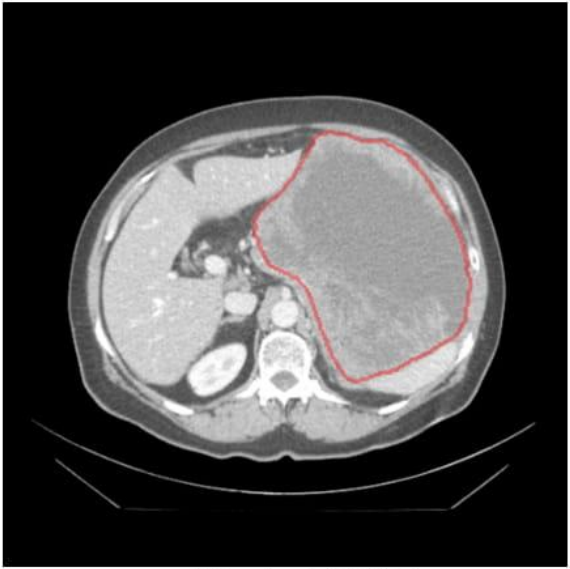
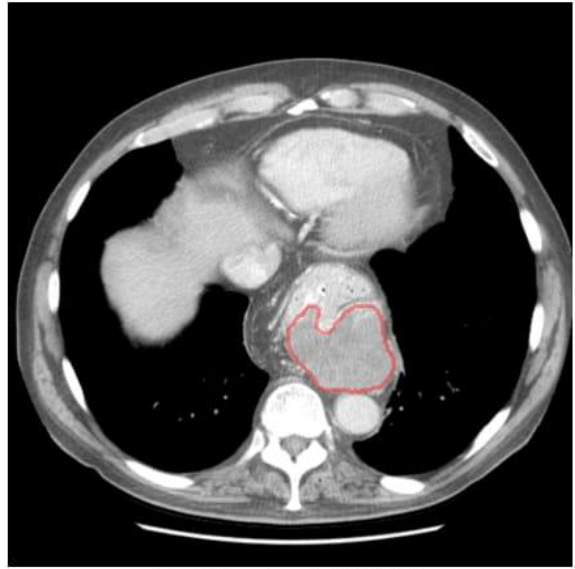
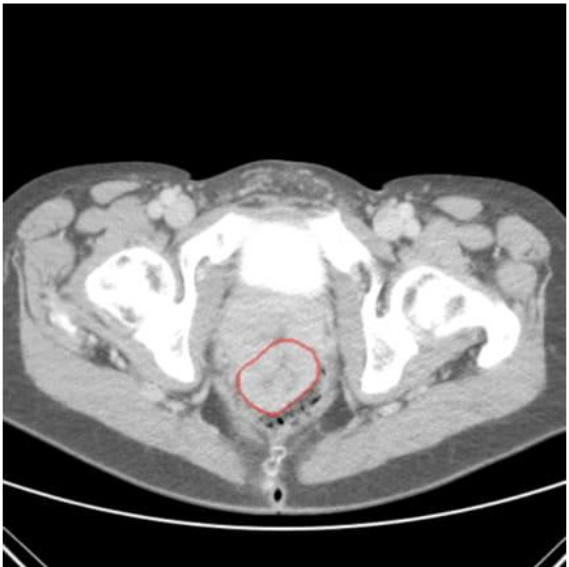
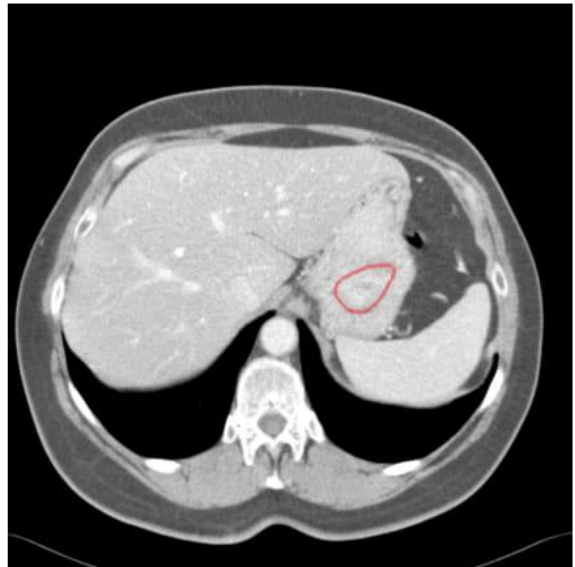

**Figure 3**. Examples of GISTs always correctly or always incorrectly predicted by the radiomics CT imaging model, i.e. model 4. The typical examples (a and b) are two of the GISTs always classified correctly by the model; the atypical examples (c and d) are two of the GISTs always classified incorrectly by the model.

**Discussion**

Radiomics can distinguish GISTs from other intra-abdominal tumors with a performance similar to three radiologists. Radiomics could not predict the presence and subtype of *c-KIT* mutations or the MI.

Diagnosing GISTs is currently done through a biopsy, aided by manually scored imaging features (Akahoshi, Oya, Koga, & Shiratsuchi, 2018; Lau et al., 2004; Liu, Liu, & Jin, 2019). The ability to distinguish GISTs from non-GISTs on routine CT scans through radiomics could be a non-invasive and quick alternative for the initial assessment of intra-abdominal tumors. The use of our model would aid quick referral of GIST patients from a peripheral hospital to a center of expertise without the need to wait for an invasive biopsy and time-consuming pathology analysis, and it would prevent GIST patients being missed (i.e. false negatives), unnecessary referral or even treatment for non-GIST (i.e. false positives). Additionally, for non-GIST benign abnormalities, our differential diagnosis model prevents further dissemination investigation and pathologic examinations. To our knowledge, this is the first study to evaluate the GIST differential diagnosis on many locations through an automated radiomics approach on a large, multi-scanner dataset and compare the performance with radiologists.

The performance of the differential diagnosis imaging-only model was similar to two radiologists, and significantly better than one. The agreement between the radiologists was poor, indicating observer dependence in the prediction, and there were significant performance differences. The advantage of the radiomics model is that it is automatic and observer independent, assuming the segmentation is reproducible as indicated by the high DSC, and that it will always give the same prediction on the same image, thereby improving over manual scoring.

Tumor location is highly relevant for distinguishing GISTs from non-GISTs as GISTs grow typically in the stomach or small intestines (Soreide et al., 2016). In our study, tumor location was based on radiology reports, which is subjective and occasionally fails to report the true tumor primary origin (Miettinen & Lasota, 2006a). Moreover, the tumor location distribution in our dataset may not be a correct representation of the overall population, e.g. only non-GISTs were located in the uterus. Despite the subjectivity of potential bias in tumor location, we added location to the imaging model for a fair comparison with the radiologists. Further research on location-matched datasets is required to investigate the value of location in the GIST differential diagnosis model.

In the literature, risk classification of outcomes such as recurrence or aggressive behavior for GISTs has mostly been based on criteria such as the Armed Forces Institute of Pathology criteria, modified National Institutes of Health consensus criteria of 2008, and the modified Fletcher classification system (Fletcher et al., 2002; Joensuu, 2008; Jones, 2014; Miettinen & Lasota, 2006b; Milliron, Mittal, Camacho, Datir, & Moreno, 2017). Several studies to evaluate radiomics for this risk stratification have been conducted over the last years (Ba-Ssalamah et al., 2013; Chen et al., 2019; Feng et al., 2018; Kurata et al., 2018; Liu et al., 2018; Ning et al., 2019; Xu, Ma, et al., 2018; Yang et al., 2018; Zhou et al., 2016; Zhuo et al., 2018). These studies illustrate the clinical need for new methods to stratify GISTs for guiding treatment decisions and show the potential of applying radiomics in the setting of GIST.

Our first contribution with respect to the existing literature is the focus on the diagnostic trajectory of GISTs by aiming to predict the differential diagnosis, whereas existing studies mainly focus on risk classification (Fletcher et al., 2002; Joensuu, 2008; Jones, 2014; Miettinen & Lasota, 2006b; Milliron et al., 2017). Second, our method determines the optimal radiomics pipeline from a large number of radiomics algorithms and parameters, automatically evaluating a large number of radiomics methods, whereas existing studies typically report the results of a ``hand-crafted'', manually optimized radiomics pipeline (Ba-Ssalamah et al., 2013; Chen et al., 2019; Feng et al., 2018; Kurata et al., 2018; Liu et al., 2018; Ning et al., 2019; Xu, Ma, et al., 2018; Yang et al., 2018; Zhou et al., 2016; Zhuo et al., 2018). Moreover, through an extensive cross-validation scheme, all model optimization was performed on the training dataset, eliminating the risk of overfitting of the model on the test set. This increases the chances of generalizability of our performance estimates. Lastly, we evaluated the model's robustness to segmentation and scanner variations. In our results, neither ICC-based features selection nor ComBat harmonization substantially altered the performance. As the model performance did not alter when using these measures to increase radiomics robustness, this may indicate that the model is already robust to these variations. Evaluating ComBat with the differential diagnosis (e.g. GIST or non-GIST) as moderation variable did lead to a near perfect performance (AUC of 0.99), but similar results were obtained with randomly labeling patients as GIST or non-GIST, indicating that this near perfect performance was a result of overfitting by ComBat.

Our study has several limitations. First, there was substantial heterogeneity in the acquisition protocols. This heterogeneity may have (negatively) affected the performance. Nevertheless, even on this heterogeneous dataset, the radiomics model achieved promising performance, similar to three experienced radiologists, suggesting high generalizability of the model. Second, the dataset for the mutation analysis was small (N=98 GISTs with known *c-KIT* mutation), which may have been too small for radiomics to learn from. Third, the use of different gene panels for the GIST mutational analysis over the years might have led to a potential underestimation of mutation prevalence in the current cohort, as newer sequencing techniques use larger gene panels and have a higher sensitivity. Additionally, only for a subset of the patients (e.g. 90 of the 125 (72%) in the MI analysis) complete histologic data was available. No data regarding the clinical outcome such as survival or recurrence was available for the GISTs. Fourth, the current radiomics approach requires manual segmentation. While accurate, this process is also time consuming and potentially subject to observer variability, although the DSC indicated good agreement and our ICC-based feature selection shows that only using reliable features resulted in a similar performance as using all features. Automatic segmentation methods, such as using deep learning, may help to overcome this limitation. Lastly, the current study has a retrospective study design. A prospective study confirming our results is needed.

Future work should focus on the external validation of our findings on an independent, external dataset. Additionally, extension of the dataset will lead to more statistical power, may improve the performance as the model has more cases to learn from, and may facilitate more data driven approaches such as deep learning. Also, this may result in sufficient samples to study prediction of *PDGFRA*, *BRAF*, and other rare *c-KIT* mutations. Importantly, an alternative to non-invasively determine the mutational status of a GIST is by ctDNA (Xu, Chen, et al., 2018). With better performance of both methods, the combination of radiomics and ctDNA assessment would allow to assess in patients with metastatic disease the most important determinants rendering an invasive biopsy redundant. Eventually, this may be followed by a prospective clinical trial with harmonized acquisition protocols in which the performance, as well as the cost-effectiveness, are assessed.

**Conclusion**

Our radiomics model was able to distinguish GIST from non-GIST intra-abdominal tumors based on pre-treatment CT imaging with a performance similar to three experienced radiologists. Our model may therefore aid clinicians early on in the diagnostic chain. The model was not able to predict the c-KIT mutational status and the MI.

**Acknowledgements**

Martijn P. A. Starmans acknowledges funding from the research program STRaTeGy (project number 14929-14930), which is (partly) financed by the Netherlands Organisation for Scientific Research (NWO). This work was partially carried out on the Dutch national e-infrastructure with the support of SURF Cooperative.

**Competing Interests**

Wiro J. Niessen is founder, scientific lead and stock holder of Quantib BV. The other authors do not declare any conflicts of interest.

**Supplemental Materials**

**Supplemental Material 1: Radiomics feature extraction**

This supplemental material is similar to (Timbergen et al., 2020; Vos et al., 2019), but details relevant for the current study are highlighted.

A total of 564 radiomics features were used in this study. All features were extracted using the defaults for CT scans from the Workflow for Optimal Radiomics Classification (WORC) (Starmans, Van der Voort, Phil, & Klein, 2018), which internally uses the PREDICT (van der Voort & Starmans, 2018) and PyRadiomics (Van Griethuysen et al., 2017) feature extraction toolboxes. An overview of all features is depicted in **Supplemental Table S3**. For details on the mathematical formulation of the features, we refer the reader (Zwanenburg et al., 2020). More details on the extracted features can be found in the documentation of the respective toolboxes, mainly the WORC documentation (Starmans, 2018).

For CT scans, the images are by default not normalized as the scans already have a fixed unit and scale (i.e. Hounsfield), contrary to MRI. The images were not resampled, as this would result in interpolation errors. The code to extract the features has been published open-source (Starmans, 2020).

The features can be divided in several groups. Thirteen intensity features were extracted using the histogram of all intensity values within the ROIs and included several first-order statistics such as the mean, standard deviation and kurtosis. These describe the distribution of Hounsfield units within the lesion. Thirty-five shape features were extracted based only on the ROI, i.e. not using the image, and included shape descriptions such as the volume, compactness and circular variance. These describe the morphological properties of the lesion. Nine orientation features were used, describing the orientation of the ROI, i.e. not using the image. Lastly, 483 texture features were extracted using Gabor filters (144 features), Laplacian of Gaussian filters (36 features), vessel (i.e. tubular structures) filters (36 features) (Frangi, Niessen, Vincken, & Viergever, 1998), the Gray Level Co-occurrence Matrix (144 features) (Zwanenburg et al., 2020), the Gray Level Size Zone Matrix (16 features) (Zwanenburg et al., 2020), the Gray Level Run Length Matrix (16 features) (Zwanenburg et al., 2020), the Gray Level Dependence Matrix (14 features) (Zwanenburg et al., 2020), the Neighbourhood Grey Tone Difference Matrix (5 features) (Zwanenburg et al., 2020), Local Binary Patterns (18 features) (Ojala, Pietikainen, & Maenpaa, 2002), and local phase filters (36 features) (Kovesi, 1997, 2003). These features describe more complex patterns within the lesion, such as heterogeneity, occurrence of blob-like structures, and presence of line patterns.

Most of the texture features include parameters to be set for the extraction. Beforehand the values of the parameters that will result in features with the highest discriminative power for the classification at hand (e.g. GIST vs non-GIST) are not known. Including these parameters in the workflow optimization, see Supplemental Material 2, would lead to repeated computation of the features, resulting in a redundant decrease in computation time. Therefore, alternatively, these features are extracted at a range of parameters as is default in WORC. The hypothesis is that the features with high discriminative power will be selected by the feature selection methods and/or the machine learning methods as described in Supplemental Material 2. The parameters used are described in **Supplemental Table S3**.

The dataset used in this study is heterogeneous in terms of acquisition protocols. Especially the variations in slice may cause feature values to be dependent on the acquisition protocol. Hence, extracting robust 3D features may be hampered by these variations, especially for low resolutions. To overcome this issue, all features were extracted per 2D axial slice and aggregated over all slices, which is default in WORC. Afterwards, several first-order statistics over the feature distributions were evaluated and used in the machine learning approach.

**Supplemental Material 2: Adaptive workflow optimization for automatic decision model creation**

This appendix is similar to (Timbergen et al., 2020; Vos et al., 2019), but details relevant for the current study are highlighted.

The Workflow for Optimal Radiomics Classification (WORC) toolbox (Starmans et al., 2018) makes use of adaptive algorithm optimization to create the optimal performing workflow from a variety of methods. WORC defines a workflow as a sequential combination of algorithms and their respective parameters. To create a workflow, WORC includes algorithms to perform feature scaling, feature imputation, feature selection, oversampling, and machine learning. If used, as some of these steps are optional as described below, these methods are performed in the same order as described in this appendix. More details can be found in the WORC documentation (Starmans, 2018). The code to use WORC for creating the differential diagnosis and molecular analysis decision models in this specific study has been published open-source (Starmans, 2020).

Feature scaling was performed to make all features have the same scale, as otherwise the machine learning methods may focus only on those features with large values. This was done through z-scoring, i.e. subtracting the mean value followed by division by the standard deviation, for each individual feature. In this way, all features had a mean of zero and a variance of one. A robust version of z-scoring was used, in which outliers, i.e. values below the 5th percentile or above the 95th percentile, were excluded from computing the mean and variance.

When a feature could be computed, e.g. a lesion is too small for specific feature to be extracted or a division by zero occurs, feature imputation was used to estimate replacement values for the missing values. Strategies for imputation included 1) the mean; 2) the median; 3) the most frequent value; and 4) a nearest neighbor approach.

Feature selection was performed to eliminate features which were not useful to distinguish between the classes, e.g. GIST vs. non-GIST. These included; 1) a variance threshold, in which features with a low variance (<0.01) are removed. This method was always used, as this serves as a feature sanity check with almost zero risk of removing relevant features; 2) optionally, a group-wise search, in which specific groups of features (i.e. intensity, shape, and the subgroups of texture features as defined in Supplemental Material 1) are selected or deleted. To this end, each feature group had an on/off variable which is randomly activated or deactivated, which were all included as hyperparameters in the optimization; 3) optionally, individual feature selection through univariate testing. To this end, for each feature, a Mann-Whitney U test was performed to test for significant differences in distribution between the labels (e.g. GIST vs non-GIST). Afterwards, only features with a p-value above a certain threshold were selected. A Mann-Whitney U test was chosen as features may not be normally distributed and the samples (i.e. patients) were independent; and 4) optionally, principal component analysis (PCA), in which either only those linear combinations of features were kept which explained 95% of the variance in the features or a limited number of components (between 10 – 50). These feature selection methods may be combined by WORC, but only in the mentioned order.

Various resampling strategies can optionally be used, which can be used to overcome class imbalances and reduce overfitting on specific training samples. These included various methods from the imbalanced-learn toolbox (Lemaitre, Nogueira, & Aridas, 2017); random over-sampling, random under-sampling, near-miss resampling, the neighborhood cleaning rule, ADASYN, and SMOTE (regular, borderline, Tomek and the edited nearest neighbors).

Lastly, machine learning methods were used to determine a decision rule to distinguish the classes. These included; 1) logistic regression; 2) support vector machines; 3) random forests; 4) naive Bayes; and 5) linear and quadratic discriminant analysis.

Most of the included methods require specific settings or parameters to be set, which may have a large impact on the performance. As these parameters have to be determined before executing the workflow, these are so-called "hyperparameters". In WORC, all parameters of all mentioned methods are treated as hyperparameters, since they may all influence the decision model creation. WORC simultaneously estimates which combination of algorithms and hyperparameters performs best. A comprehensive overview of all parameters is provided in the WORC documentation (Starmans, 2018).

By default in WORC, the performance is evaluated in a 100x random-split train-test cross-validation. In the training phase, a total of 25,000 pseudo-randomly generated workflows is created. These workflows are evaluated in a 5x random-split cross-validation on the training dataset, using 85% of the data for actual training and 15% for validation of the performance. All described methods are fit on the training datasets, and only tested on the validation datasets. The workflows are ranked from best to worst based on their mean performance on the validation sets using the F1-score, which is the harmonic average of precision and recall. Due to the large number of workflows that is executed, there is a chance that the best performing workflow is overfitting, i.e. looking at too much detail or even noise in the training dataset. Hence, to create a more robust model and boost performance, WORC combines the 50 best performing workflows into a single decision model, which is known as ensembling. These 50 best performing workflows are re-trained using the entire training dataset, and only tested on the test datasets. The ensemble is created through averaging of the probabilities, i.e. the chance of a patient being GIST or non-GIST, of these 50 workflows.

A full experiment consists of executing 12.5 million workflows (25,000 pseudo-randomly generated workflows, times a 5x train-validation cross-validation times 100x train-test cross-validation), which can be parallelized. The computation time of training or testing a single workflow is on average less than a second, depending on the size of the dataset both in terms of samples (i.e. patients) and features. The largest experiment in this study, i.e. the differential diagnosis including 247 patients had a computation time of approximately 32 hours on a 32 CPU core machine. The contribution of the feature extraction to the computation time was negligible.

**Supplemental References**

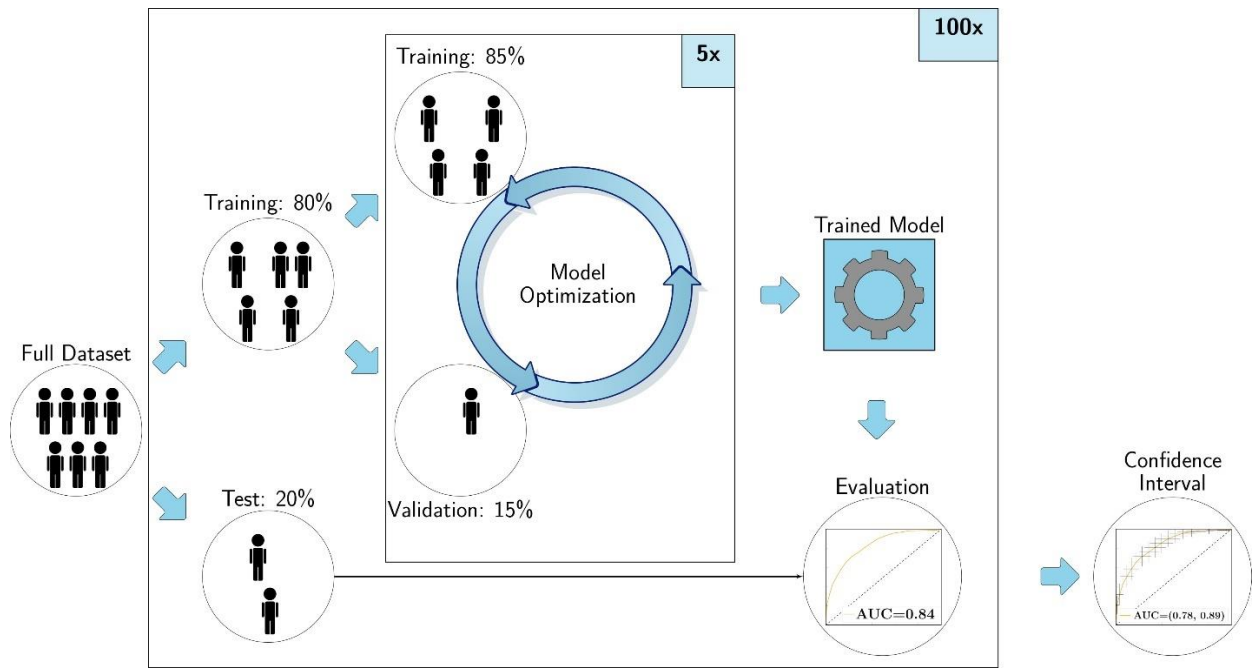

**Supplemental Figure S1.** Visualization of the 100x random split-cross validation, including a second cross validation within the training set.

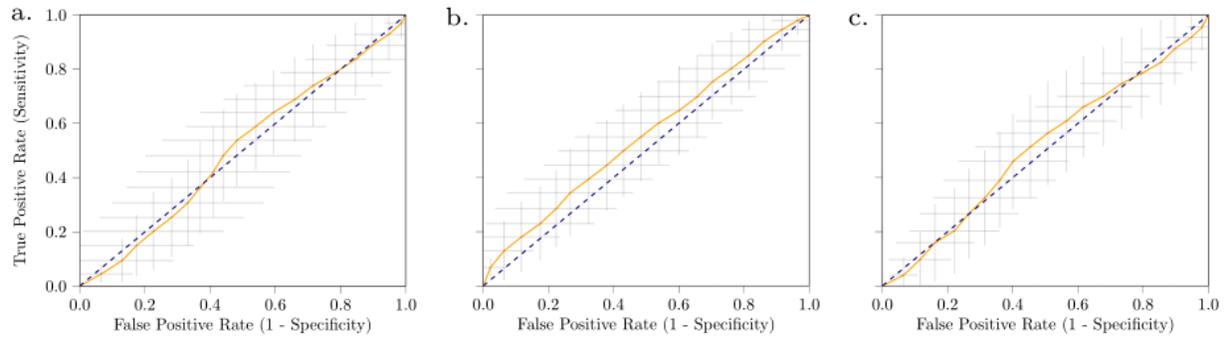

**Supplemental Figure S2.** Receiver operating characteristic curves of the radiomics models based on the CT imaging features, age at diagnosis and sex for (**a**) *c-KIT* presence vs. absence; (**b**) *c-KIT* exon 11 presence vs. absence; and (**c**) mitotic index (≤5/50 HPF vs. >5/50 HPF). The grey crosses identify the 95% confidence intervals of the 100x random-split cross-validation; the orange curve is fit through their means.

**Supplemental Table S1.** P-values of features from univariate tests between GIST and non-GIST patients after Bonferonni correction. A Mann-Whitney U test was used for continuous variables, a Chi-square test for categorical variables. Only features with a p-value < 0.05, which are considered statistically significant, are shown. Besides the feature names, several of the feature labels also include the parameters used. More details on the features can be found in Supplemental Materials 1.

| Label | Mann-Whitney U P | Chi2 P |
|---|---|---|
| semf_location | | 1.14328E-17 |
| of_COM_x | 7.45991E-08 | |
| hf_energy | 1.59291E-05 | |
| of_COM_Index_x | 3.05687E-05 | |
| tf_Gabor_mean_F0.2_A0.79 | 0.00033747 | |
| tf_GLRLM_LongRunEmphasis | 0.001104272 | |
| tf_GLRLM_RunVariance | 0.001152602 | |
| tf_GLSZM_ZonePercentage | 0.001298949 | |
| tf_GLRLM_ShortRunEmphasis | 0.001310058 | |
| tf_GLRLM_RunPercentage | 0.001366976 | |
| tf_GLRLM_RunLengthNonUniformityNormalized | 0.001366976 | |
| tf_GLDM_DependenceVariance | 0.001488 | |
| tf_Gabor_mean_F0.2_A2.36 | 0.001500651 | |
| tf_GLDM_LargeDependenceEmphasis | 0.001578749 | |
| tf_GLDM_SmallDependenceLowGrayLevelEmphasis | 0.002973555 | |
| tf_GLCMMS_homogeneityd3.0A0.79mean | 0.004120245 | |
| tf_Gabor_energy_F0.2_A0.79 | 0.004919574 | |
| tf_GLRLM_LongRunHighGrayLevelEmphasis | 0.005865334 | |
| tf_GLDM_SmallDependenceEmphasis | 0.006451746 | |
| tf_Gabor_energy_F0.2_A2.36 | 0.007206412 | |
| sf_area_std_2D | 0.007206412 | |
| tf_GLDM_DependenceNonUniformityNormalized | 0.008365756 | |
| tf_GLDM_LargeDependenceLowGrayLevelEmphasis | 0.008699116 | |
| tf_Gabor_energy_F0.2_A1.57 | 0.008974852 | |
| tf_GLDM_LargeDependenceHighGrayLevelEmphasis | 0.010242097 | |
| tf_Gabor_mean_F0.2_A0.0 | 0.011411483 | |
| tf_GLCMMS_homogeneityd3.0A0.0mean | 0.012609864 | |
| sf_area_max_2D | 0.01480121 | |
| tf_GLSZM_LargeAreaHighGrayLevelEmphasis | 0.015967243 | |
| tf_GLSZM_LargeAreaEmphasis | 0.01621066 | |
| tf_GLSZM_ZoneVariance | 0.016457589 | |
| tf_Gabor_energy_F0.2_A0.0 | 0.016582385 | |
| hf_min | 0.017060365 | |
| sf_area_avg_2D | 0.022211237 | |
| tf_GLRLM_LongRunLowGrayLevelEmphasis | 0.024458935 | |
| of_COM_y | 0.02537934 | |
| tf_GLSZM_LargeAreaLowGrayLevelEmphasis | 0.026722996 | |
| tf_Gabor_median_F0.05_A2.36 | 0.0270312 | |
| sf_shape_Maximum2DDiameterSlice | 0.031168368 | |
| tf_GLRLM_GrayLevelNonUniformityNormalized | 0.038741953 | |
| vf_Frangi_inner_energy_SR(1.0. 10.0)_SS2.0 | 0.039022054 | |
| tf_Gabor_mean_F0.5_A2.36 | 0.04472035 | |
| tf_Gabor_kurtosis_F0.05_A0.79 | 0.045688679 | |

*Abbreviations: GLCM: gray level co-occurrence matrix; GLCMMS: GLCM multislice; NGTDM: neighborhood gray tone difference matrix; GLSZM: gray level size zone matrix; GLRLM: gray level run length matrix; LBP: local binary patterns; LoG: Laplacian of Gaussian; std: standard deviation .

**Supplemental Table S2.** Performance of the radiomics models for the differential diagnosis based on imaging using only features with good (ICC > 0.75) or excellent (ICC > 0.90) reliability; and using ComBat harmonization per manufacturer or per protocol (manufacturer and high/low slice thickness). For each metric, the mean and 95% confidence interval over the 100x random-split cross-validation iterations are given.

|  | ICC > 0.75 | ICC > 0.90 | ComBat Manufacturer | ComBat Protocol |
|---|---|---|---|---|
| **AUC** | 0.76 [0.70, 0.82] | 0.75 [0.69, 0.82] | 0.76 [0.69, 0.82] | 0.73 [0.66, 0.80] |
| **BCA** | 0.70 [0.64, 0.76] | 0.69 [0.63, 0.75] | 0.70 [0.65, 0.75] | 0.67 [0.61, 0.73] |
| **Sensitivity** | 0.64 [0.52, 0.76] | 0.59 [0.49, 0.70] | 0.65 [0.56, 0.74] | 0.60 [0.49, 0.71] |
| **Specificity** | 0.75 [0.66, 0.84] | 0.79 [0.71, 0.88] | 0.75 [0.65, 0.85] | 0.75 [0.66, 0.84] |

* Abbreviations: AUC: area under the receiver operating characteristic curve; BCA: balanced classification accuracy.

**Supplemental Table S3.** Overview of the 564 features used in this study. GLCM features were calculated in four different directions (0, 45, 90, 135 degrees) using 16 gray levels and pixel distances of 1 and 3. LBP features were calculated using the following three parameter combinations: 1 pixel radius and 8 neighbours, 2 pixel radius and 12 neighbours, and 3 pixel radius and 16 neighbours. Gabor features were calculated using three different frequencies (0.05, 0.2, 0.5) and four different angles (0, 45, 90, 135 degrees). LoG features were calculated using three different widths of the Gaussian (1, 5 and 10 pixels). Vessel features were calculated using the full mask, the edge, and the inner region. Local phase features were calculated on the monogenic phase, phase congruency and phase symmetry.

| Histogram (13 features) | LoG (13*3=39 features) | Vessel (12*3=39 features) | GLCM (MS) (6*3*4*2=144 features) | Gabor (13*4*3=156 features) | NGTDM (5 features) | LBP (13*3=39 features) |
|---|---|---|---|---|---|---|
| min | min | min | contrast (normal, MS mean + std) | min | busyness | min |
| max | max | max | dissimilarity (normal, MS mean + std) | max | coarseness | max |
| mean | mean | mean | homogeneity(normal, MS mean + std) | mean | complexity | mean |
| median | median | median | angular second moment (ASM) (normal, MS mean + std) | median | contrast | median |
| std | std | std | energy (normal, MS mean + std) | std | strength | std |
| skewness | skewness | skewness | correlation (normal, MS mean + std) | skewness | | skewness |
| kurtosis | kurtosis | kurtosis | | kurtosis | | kurtosis |
| peak | peak | peak | | peak | | peak |
| peak position | peak position | peak position | | peak position | | peak position |
| range | range | range | | range | | range |
| energy | energy | energy | | energy | | energy |
| quartile range | quartile | quartile | | quartile range | | quartile range |
| entropy | entropy | entropy | | entropy | | entropy |

| GLSZM (16 features) | GLRM (16 features) | GLDM (14 features) | Shape (35 features) | Orientation (9 features) | Local phase (13*3=39 features) |
|---|---|---|---|---|---|
| Gray Level Non Uniformity | Gray Level Non Uniformity | Dependence Entropy | compactness (mean + std) | theta_x | min |
| Gray Level Non Uniformity Normalized | Gray Level Non Uniformity Normalized | Dependence Non-Uniformity | radial distance (mean + std) | theta_y | max |
| Gray Level Variance | Gray Level Variance | Dependence Non-Uniformity Normalized | roughness (mean + std) | theta_z | mean |
| High Gray Level Zone Emphasis | High Gray Level Run Emphasis | | convexity (mean + std) | COM index x | median |
| Large Area Emphasis | Long Run Emphasis | Dependence Variance | circular variance (mean + std) | COM index y | std |
| Large Area High Gray Level Emphasis | Long Run High Gray Level Emphasis | Gray Level Non-Uniformity | principal axes ratio (mean + std) | COM index z | skewness |
| Large Area Low Gray Level Emphasis | Long Run Low Gray Level Emphasis | Gray Level Variance | elliptic variance (mean + std) | COM x | kurtosis |
| Low Gray Level Zone Emphasis | Low Gray Level Run Emphasis | High Gray Level Emphasis | solidity (mean + std) | COM y | peak |
| SizeZoneNonUniformity | RunEntropy | Large Dependence Emphasis | area (mean, std, min + max | COM z | peak position |
| SizeZoneNonUniformityNormalized | RunLengthNonUniformity | Large Dependence High Gray Level Emphasis | volume (total, mesh, volume) | | range |
| SmallAreaEmphasis | RunLengthNonUniformityNormalized | | elongation | | energy |
| SmallAreaHighGrayLevelEmphasis | RunPercentage | Large Dependence Low Gray Level Emphasis | flatness | | quartile |
| SmallAreaLowGrayLevelEmphasis | RunVariance | Low Gray Level Emphasis | least axis length | | entropy |
| ZoneEntropy | ShortRunEmphasis | Small Dependence Emphasis | major axis length | | |
| ZonePercentage | ShortRunHighGrayLevelEmphasis | Small Dependence High Gray Level Emphasis | minor axis length | | |
| ZoneVariance | ShortRunLowGrayLevelEmphasis | Small Dependence Low Gray Level Emphasis | maximum diameter 3D | | |
| | | | maximum diameter 2D (rows, columns, slices) | | |
| | | | sphericity | | |
| | | | surface area | | |
| | | | surface volume ratio | | |

*Abbreviations: COM: center of mass; GLCM: gray level co-occurrence matrix; MS: multi slice; NGTDM: neighborhood gray tone difference matrix; GLSZM: gray level size zone matrix; GLRLM: gray level run length matrix; LBP: local binary patterns; LoG: Laplacian of Gaussian; std: standard deviation.